\documentclass[preprint,aps,prl,groupedaddress,amssymb,showpacs]{revtex4}
\usepackage{graphicx}
\usepackage{dcolumn}
\usepackage{amsmath}
\begin{document}

\title{   Image of Veselago lens based upon two-dimensional photonic 
crystal with triangular lattice. }

\author{C. Y. Li}
\affiliation{University of Utah, Salt Lake City UT, 84112 USA}

\author{J. M. Holt}
\affiliation{University of Utah, Salt Lake City UT, 84112 USA}

\author{A. L. Efros}
\email{efros@physics.utah.edu}
\affiliation{University of Utah, Salt Lake City UT, 84112 USA}

\date{\today}

\begin{abstract}
The construction of the multi-focal  Veselago lens predicted earlier is proposed on the basis of a uniaxial photonic crystal consisting of cylindrical air holes in silicon that make a triangular lattice in a plane perpendicular to the axis of the crystal. The object and image are in  air. The period of the crystal should be $0.44\mu{\rm m}$ to work at the wavelength $1.5\mu{\rm m}$. The lens does not provide superlensing but the half-width of the image is $0.5\lambda$. The lens is shown to have wave guiding properties depending on the substrate material. 
\end{abstract}
\maketitle

\section{1.Introduction}
 A left handed medium (LHM), defined by Veselago\cite{ves} as a medium with simultaneously negative $\mu$ and $\epsilon$, has recently attracted much attention mostly because of negative refraction at its interface with a regular medium (RM). This effect allows creation of a unique device called the Veselago lens. This lens is a slab of a LHM embedded inside a RM with the condition that  both media have the same {\em isotropic} refractive index and the same impedance.

It was shown recently\cite{ef,pok} that a two-dimensional (2D) dielectric uniaxial photonic crystal (PC) made of non-magnetic materials can behave as a LHM with negative $\epsilon$ and $\mu$ if it has a negative group velocity in the vicinity of the $\Gamma$-point of the second Brillouin zone. Experimental demonstration of negative refraction in a metallic PC using the modes near the $\Gamma$-point has been presented in Ref. \cite{parimis}. 

Proximity to the $\Gamma$-point is crucial for the PC-based LHM. In general, $\epsilon$ and $\mu$ of a PC are functions not only of $\omega$, but also of ${\bf k}$. It has been shown\cite{ef} that, in the vicinity of the $\Gamma$-point, $\epsilon$, $\mu$, and $\omega$ are functions of $k$. Therefore, for every propagating mode, ${\bf k}$, $\epsilon$, and $\mu$ can be represented as a function of $\omega$. However, the parameters $\epsilon(\omega)$ and $\mu(\omega)$ are a property of a given mode in the medium rather than a property of the whole medium. It is shown\cite{cond1,cond2} that parameters $\epsilon$ and $\mu$ for evanescent waves (EW's) in the same medium and at the same frequency strongly depend on ${\bf k}$. 

We think now that the 2D PC with a square lattice considered in Ref. \cite{ef,cond1,cond2} is a wrong choice from a practical point of view. Although both $\mu$ and refractive index $n$ are zero\cite{ef} exactly at the $\Gamma$-point, one should still work near the $\Gamma$-point where the parameters $\epsilon$, $\mu$, and $\omega$ are isotropic. In the square lattice the maximum value of $n$ that can be achieved in the isotropic region is about 0.33. For the Veselago lens, the LHM and the RM should have the same value of $n$ and it would be difficult to find 
a transparent RM with such a low $n$.
The situation is much better for the 2D lattice with hexagonal symmetry in the plane (triangular lattice shown in Fig. 1(a)) because the width of the isotropic region near the $\Gamma$-point is much wider. Wang {\it et al.}\cite{w1} were the first to find out computationally that $n=1$ can be reached in the isotropic region of the hexagonal (triangular) lattice. Their result can be explained from symmetry. For both square and triangular lattices, the expansions of frequency near the $\Gamma$-point have the same form up to the quadratic term, namely, $\omega^2= \omega_0^2-\alpha k^2$. In this approximation, the frequency of the mode is isotropic. The question is, what is the order of magnitude of the first anisotropic term in this expansion? Group theory provides a simple answer. In a square lattice with the symmetry axis $C_4$, the largest anisotropic term is of the order of $k^4$ and has a form $k_{x}^2k_{y}^2$. In a triangular lattice with axis $C_6$, this term is of the order of $k^6$ and has a more complicated structure. Therefore the isotropic region is much wider in the latter case. Common sense also suggests that the anisotropy should be less in the lattice with six nearest neighbors than in the lattice with four neighbors.

\begin{figure}
\includegraphics[width=8.6cm]{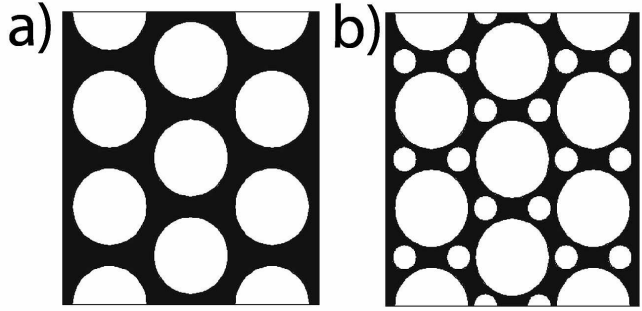}
\caption{(a) Triangular lattice of circular cylindrical holes in a dielectric matrix with $\epsilon_m=12$, $\mu_m=1$. The radius of the holes $R=0.35d$, where $d$ is the period. (b) The
 lattice with the basis and parameters as in Ref. 8. The radii of large and small air holes are $R=0.4d$ and $R=0.13d$, respectively. The dielectric matrix has $\epsilon_m = 12.96$, $\mu_m=1$.
 \label{fig1}}
\end{figure}
  
With this finding, the creation of the Veselago lens based upon photonic crystal at the frequency of long-distance communication (1.5$\mu{\rm m}$) becomes a reality. However, even if we are able to match refractive index with air, effective $|\epsilon|$ and $|\mu|$ of the PC would still be different from those in air. In this case, the impedances are not matched, reflection is not zero, and one gets a multi-focal lens considered in Ref. \cite{poef}. The advantage of the multi-focal lens is that its object domain is not limited to a distance $L$ from the slab, as in the case of a matched Veselago lens. Instead, it is just a half-space in front of the infinite LHM slab. The disadvantage is that part of the energy flux gets lost due to reflection.

In this paper we simulate the imaging of a point source by a uniaxial PC with the triangular lattice in a plane perpendicular to the axis of the PC. We demonstrate the focus in the case $a>L$, where $a$ is the distance from the point source to the lens. In the case $a<L$, we find two foci behind the slab that are located in predicted places.

Wang {\it et al.}\cite{w1,w2} also considered theoretically focusing in a similar slab matched with air. They claim observation of ``superlensing'' or even ``unrestricted superlensing.'' We have found ``regular'' lensing rather than ``superlensing'' and explain it by decay of the evanescent waves (EW's) in the PC. The same authors\cite{w2} found out that by adding a basis to the lattice one can improve ``superlensing'' in the far-field regime. The best choice for improvement is shown in Fig. 1(b). This improvement strongly contradicts
 to the theory of the PC-based LHM\cite{cond1,cond2}. This theory predicts that in the far-field region the intensity of the image for the infinite lens depends on the wavelength but, since the wavelength is larger than the lattice period $d$, it does not depend on the crystalline structure. We checked the statement of Wang {\it et al.}\cite{w2} and have not found any improvement of the image for the PC with basis.

Our paper is organized as follows. In section 2 we present all computational results concerning the image of the Veselago lens based upon the PC with a triangular structure with and without basis and compare these results with a previous theory of regular lensing. In section 3 we carefully study the behavior of the EW's in this PC. In sections 2-3 we assume that the PC is infinite in the direction of its axis. This is not realistic since to work at 1.5$\mu{\rm m}$ we need a PC with period about 0.4$\mu{\rm m}$. Even with current silicon technology one should not expect to get deep holes at such a scale. Therefore in section 4 we consider the propagation of waves through the PC slab with the depth of the holes less than the wavelength. We find that the slab exhibits a strong wave guiding effect at the working frequency and propose an explanation for it.

\section{2. Images of the multi-focal Veselago lens}
In sections 2,3 we use  for our simulation a uniaxial PC infinite in the direction of the axis of the crystal ($z$-axis). Thus we have a 2D problem in the $x-y$ plane. The cross-section of the PC by this plane is shown in Fig. 1. To begin, we discuss the lattice without basis (Fig. 1(a)). All our simulations were built using the finite-element software package FEMLAB. Fig. 2 shows the spectrum of s-polarized propagating waves ($H_z=0$) for this PC.

\begin{figure}
\includegraphics[width=8.6cm]{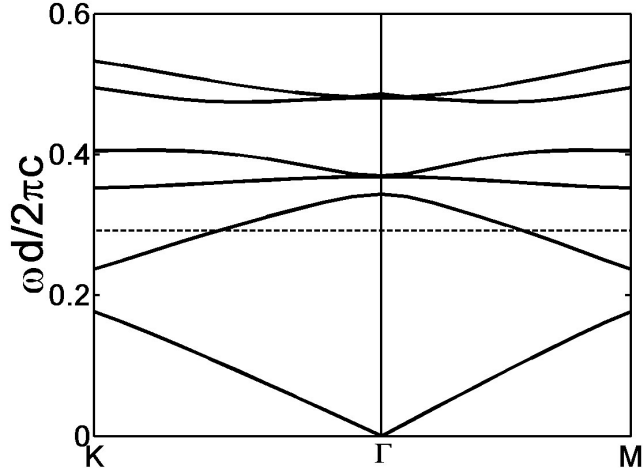}
\caption{  Six lowest bands of the photonic spectrum of the PC that has the geometry of Fig. 1(a). 
 \label{fig2}}
\end{figure}

The dashed line shows our working frequency $\omega_w d/2\pi c=0.292$, where $d$ is the period of the crystal. At this frequency the refractive index $n$, as defined by equation $n = ck/\omega$, is 1 and the wavelength $\lambda=d/0.292$.
Therefore one can get  wavelength $\lambda=1.5\mu{\rm m}$ under the condition that 
  the period $d=0.44\mu{\rm m}$ and  hole radius is $0.35 d$.
 We have checked that the spectrum is isotropic at this frequency, which means that the equation $\omega(k_x,k_y)=\omega_w$ is fulfilled 
on a curve very close to a circle in $k_x,k_y$ space. Since the group velocity at this frequency is negative, it follows from Ref. \cite{ef} that both $\epsilon$ and $\mu$ should also be isotropic and negative. However, they should not be equal to $-1$. The physical reasons for the appearance of a magnetization in the PC made of non-magnetic materials are discussed in Ref. \cite{ef}. Thus, at the working frequency, this PC is a LHM. 

We present a computer simulation of a lens made of the slab of uniaxial PC with a planar $(x,y)$ structure shown in Fig. 1(a). The slab is surrounded by a homogeneous medium with $\epsilon=\mu=1$. It is infinite in the $z$-direction, has a width $L=3.83\lambda$ in the $x$-direction and a length $H=17.52\lambda$ in the $y$-direction. A 2D point source is located on the $x$-axis at a distance $a$
in front of the slab. Thus, the problem is two-dimensional.

Our results are shown in Figs. 3 and 4. In both figures part (a) shows the distribution of electric energy in the $x-y$ plane behind the lens while part (b) shows the same distribution as a function of $x/\lambda$ at $y=0$.

\begin{figure}
\includegraphics[width=8.6cm]{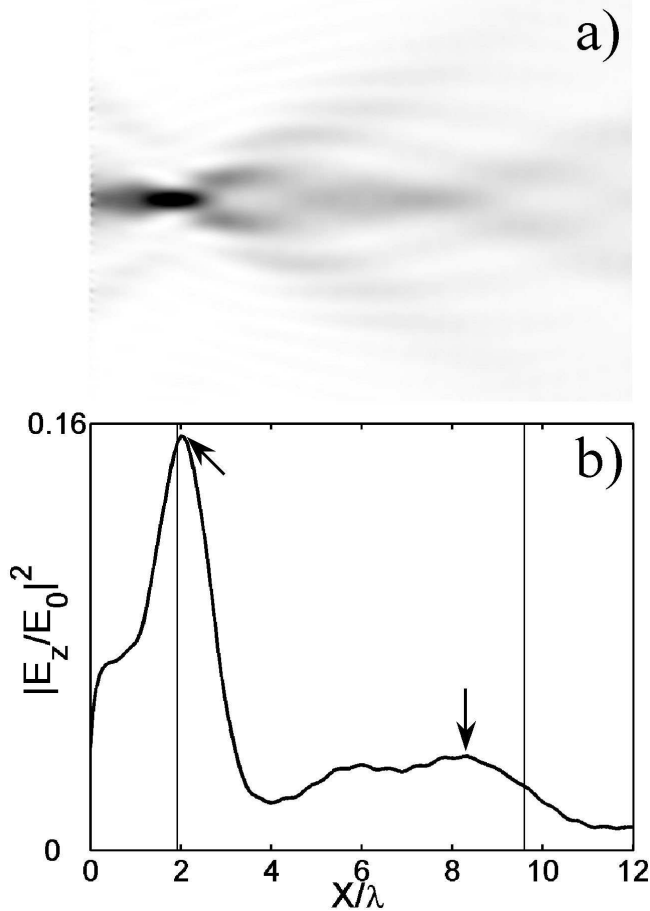}
\caption{  Distribution of electric energy  behind the PC slab for $a = 0.5 L$ in the $x,y$ plane (a) and along the line $y=0$ (b).  Two thin lines show the theoretical position of the two foci. Small arrows show computational positions of the foci.
 \label{fig3}}
\end{figure}

\begin{figure}
\includegraphics[width=8.6cm]{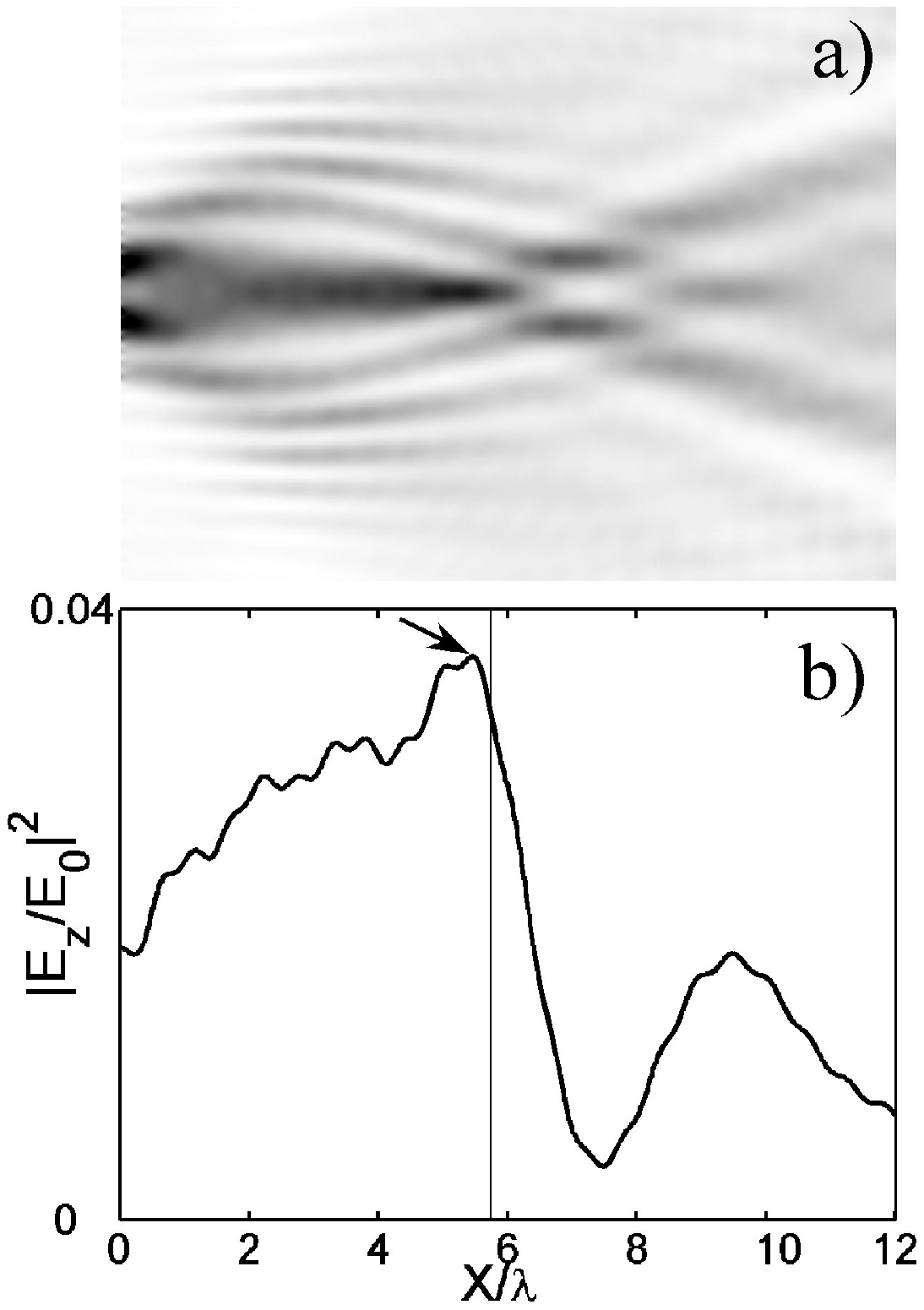}
\caption {Same as in Fig. 4  for $a = 1.5 L$. In this case we get only one focus. The second maximum is interpreted as a satellite of the focus.
 \label{fig4}}
\end{figure}

The following theoretical results are necessary to interpret these data. It is shown in Ref. \cite{poef} that the positions of the foci behind the LHM slab are described by the equation $L(2m-1)-a$ with the origin at the right surface of the slab. In the case of $a<L$, $m = 1,2,3,...$ and in the case of $a>L$, $m=2,3,4,...$. The intensity of light in any focus is given by the equation
 \begin{equation}
   I_m = (1-r^2)^2r^{4m-4}I_0.
   \label{int} 
 \end{equation} 
Here $I_0$ is the intensity of the source and $r$ is the reflection
 coefficient at a single RM-LHM interface. If $n=1$, the reflection coefficient is related to the electric permittivity as\cite{jac}
 \begin{equation}
   r=\frac{|\epsilon'|-\epsilon}{|\epsilon'|+\epsilon}
   \label{r}
 \end{equation}
and is independent of incident angle. It is important to mention that the intensity of the second focus at $a<L$ is the same as the intensity of the first focus at $a>L$. Theory also predicts the appearance of foci in front of the slab that are created by reflected light, but we do not distinguish them because of the strong intensity of the source.

The electric field in the $z$-direction produced by the point source is $E_s=iE_0H_0^{(1)}(\rho k_0)\exp{-i\omega t}$, where $k_0=\omega /c$, $n=1$, $\rho$ is a polar coordinate with the origin at the source, and the Hankel function $H^{(1)}_0=J_0+iN_0$. The Fourier transform of the source field contains both the propagating and evanescent waves\cite{cond1}. In the next section we show that the PC under study does not provide any amplification of EW's. Then we can consider only propagating waves in the Fourier transform to get the following expression\cite{cond2} for the field near any focal point $m$:
 
\begin{equation}
E_m(x',y)=\frac{iE_m}{\pi}\int_{-k_H}^{k_H} \frac{\exp i\left(k y+x^{\prime}\sqrt{k_0^2-k^2}-\omega t\right)}{\sqrt{k_0^2-k^2}}dk.
\label{m}
\end{equation}

The amplitudes $E_m<E_0$ take into account reflection and they can be related to the intensities $I_m$ given by Eq. (\ref{int}) while $x^{\prime}$ is the $x$-coordinate with the origin at a focal point $m$. The wave vector $k_H\rightarrow k_0$ when the length of the PC slab $H\rightarrow \infty$. For the finite lens
 $k_H=k_0(H/2a)/\sqrt{1+(H/2a)^2}$ (see Ref. \cite{cond1}). The energy distribution given by $|E_m(x',y)|^2$ provides the diffraction limit for the Veselago lens.

To interpret the computational results shown in Figs. 3 and 4, one should take into account that the field near the foci as given by Eq. (\ref{m}) has a main maximum at the focal point and infinite number of smaller maxima. It is also important to notice that due to reflections the amplitude $E_m$ strongly decreases with increasing $m$. Fig. 3 ($a<L$) shows the first maximum ($m=1$) exactly at the theoretical position and a long plateau with two small maxima. We think that the first of these two is a satellite of the strong maximum with $m=1$ while the second one corresponds to $m=2$. It has a small shift to the left from its theoretical position.

\begin{figure}
\includegraphics[width=8.6cm]{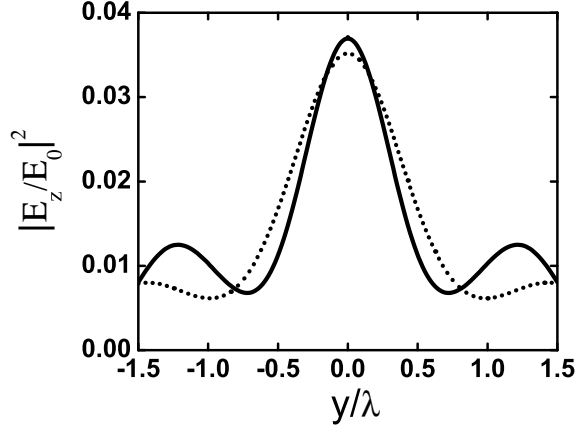}
\caption{ Solid line shows the computational distribution near the first focus for the case of $a>L$ and dotted line shows the computational distribution near the second focus for the case of $a<L$. There is no any numerical fitting in this plot.
 \label{fig5}}
\end{figure}

\begin{figure}
\includegraphics[width=8.6cm]{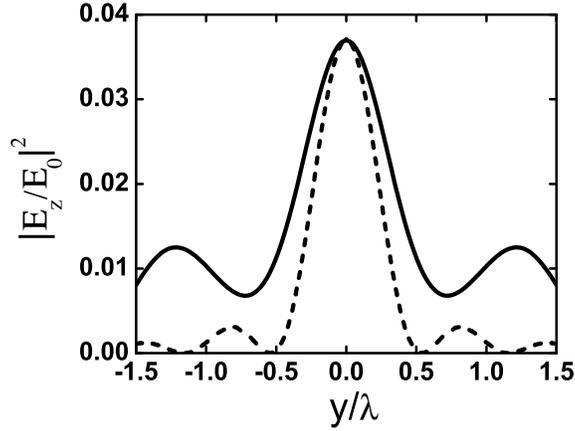}
\caption{ Distribution of electric energy along the lateral direction near the focus for the case of $a>L$. Solid line shows the computational distribution near the first focus and dashed line show the analytical result for intensity obtained from Eq. (\ref{m}). The results are normalized to have the same maximum value at the focal point.
 \label{fig6}}
\end{figure}

\begin{figure}
\includegraphics[width=8.6cm]{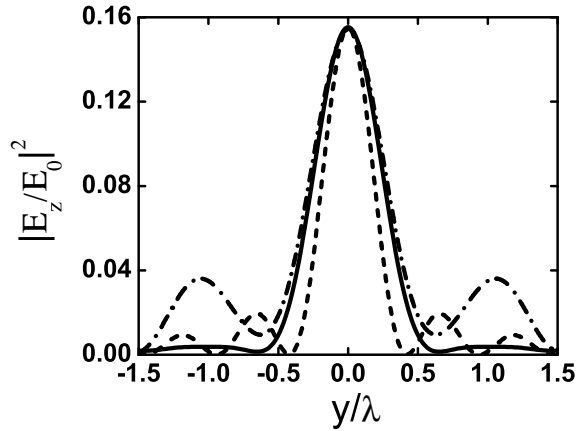}
\caption{ Distribution of electric energy along the lateral direction near the first focus for the case of $a<L$. Solid line shows the computational distribution for PC as shown in Fig. \ref{fig1}(a), dash-dotted line shows the computational distribution for PC with ``basis'' as shown in Fig. \ref{fig1}(b) and the dashed line shows the analytical result for intensity obtained from Eq. (\ref{m}). The results are normalized to have the same maximum value at the focal point for PC without ``basis''.
 \label{fig7}}
\end{figure}

In the case $a>L$, the matched Veselago lens does not produce any foci. The foci appear due to reflection and the first focus should be at a distance $3L-a$ from the lens. The first maximum in  Fig. 4  is very close to the theoretical position and we think it is the focus. We interpret the second maximum in Fig. 4 as a satellite of this focus.

 The following arguments confirm such an interpretation for $a<L$ and $a>L$:\\
a) The intensity of the focus at $a>L$ is very close to the intensity of the second focus at $a<L$. Eq. (\ref{int}) predicts that the intensities should be equal because both additional foci are formed by equal amount of reflections.\\
b) Fig. 5
 shows the lateral ($y$-direction) distribution of energy for both additional foci. One can see that the main maxima have almost the same shape. Note that no fitting parameters are used in Fig. 5.\\ 
c) In Fig. 6, the computational distribution of energy near the additional focus at $a>L$ in the lateral direction is compared with the analytical expression given by Eq. (\ref{m}). The value of $E_m$ at $m=2$ is chosen in such a way that the maximum values of both distributions are the same. The main maximum of the distribution is described satisfactorily by this equation.\\
d) We believe that the second maxima in both Fig. 3(b) and Fig. 4(b) are satellites of the previous foci rather than additional foci because they are located at the same distances from the previous foci. The minima on both curves are also located at the same distances from the previous foci. This similarity indicates that the first and the second maxima are described by the same universal curve, as it should be if the second maximum is a satellite of the first one.

Now we describe the intensity distribution in the vicinity of the main focus of the Veselago lens that exists at $a<L$ only. The lateral distribution of energy is shown in Fig. 7. One can see that the diffraction limit given by Eq. (\ref{m}) is a little narrower than the peak obtained by simulation. Thus, we have regular lensing rather than superlensing. The main peak obtained on the lattice with a basis is indistinguishable from the result without a basis. 

The simulation gives much sharper peaks in the lateral direction than in the perpendicular direction. This is a general property of Eq. (\ref{m}) that has been observed in the case of the completely matched lens as well\cite{cond1,cond2}. Note also that in the lateral direction the simulation does not reproduce the satellites.

We think that the difference between computational and analytical distributions of the satellite maxima in the lateral direction that we see in Figs. 5, 6, and 7 is due to some angular dependence of the reflection coefficient $r$ that is absent in the theory (see Eq. (\ref{r})). This dependence may appear because our working frequency is not close enough to the $\Gamma$-point and the PC cannot be perfectly described as a continuous medium. It would certainly affect the fine structure of the image more strongly. Note that the simulation describes the satellites much better in the case when reflection is small\cite{cond1,cond2}.

\section{3. Evanescent waves in triangular PC}

The initial idea of superlensing was proposed by Pendry\cite{pen} It is based upon amplification of evanescent waves by the LHM slab that should help to restore the image. Following Veselago\cite{ves}, Pendry considered a hypothetical LHM (HLHM), which is a homogeneous material with simultaneously negative $\epsilon$ and $\mu$. This amplification is shown to be related to surface polaritons\cite{hal} studied by Ruppin\cite{rup} in the HLHM. The polaritons are closely connected with negative $\epsilon$ and $\mu$ in the bulk of the HLHM.

It has been shown recently\cite{cond1,cond2} that a two-dimensional PC with a square lattice generally does not amplify EW's. However, some amplification in the near-field regime may or may not appear depending on the properties of the surface. For example, if the surface is cut between the holes (BH), amplification does not appear at all, while if it is cut across the holes (AH), amplification appears and provides superlensing in the near-field regime. The authors argue that the amplification is due to surface waves that has nothing in common with the left-handed properties of the PC with respect to propagating waves.

We show here that a 2D PC with the triangular lattice does not amplify EW's for both BH and AH cuts of the surface. The geometry of the simulation is shown 
in  Fig. 8 and the working frequency is the same as in the previous section.
The PC slab is surrounded by a homogeneous medium with $\epsilon=\mu=1$. An EW in a homogeneous medium has a form $e_z\sim \exp(ik_y y-\kappa x)$ with the dispersion law $k_y^2-\kappa^2=\omega^2/c^2$. We create such a wave in our simulation using boundary conditions for an incident wave at $x=0$ in a form $e_z = \exp{(i k_y y)}$ and $d e_z / dx = -\kappa e_z$. Note that due to reflection the total field shown in Fig. 8 may not coincide at $x=0$ with the field of the incident wave. The coordinate of the left surface of the PC slab is $x=L/2$.

\begin{figure}
\includegraphics[width=8.6cm]{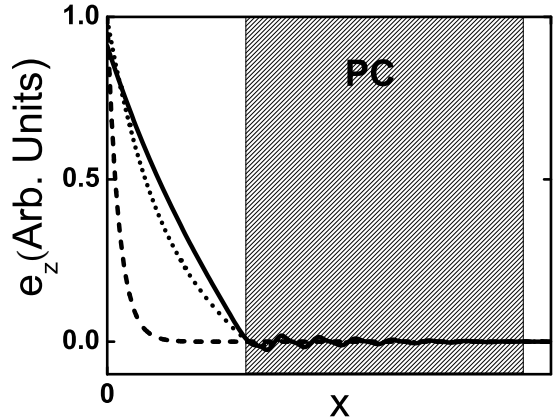}
\caption{ Electric field of the evanescent wave. The cross-section $y = 0$ is shown. The solid line is for $\kappa=0.1k_0$ and $k_y=\sqrt{1.01}k_0$, the dotted line is for $\kappa=0.2k_0$ and $k_y=\sqrt{1.04}k_0$, and the dashed line is for $\kappa=k_0$ and $k_y=\sqrt{2}k_0$. For other $\kappa$ in the interval $0.2k_0$ and $k_0$, the plots are between the dotted line and dashed line. The surface cut is BH. The plot for AH cut is indistinguishable from this plot.
 \label{fig8}}
\end{figure}

The result shows that the EW decays inside the PC slab without any signature of amplification. The surface of the PC slab is BH. We have checked that the PC slab with the AH surface gives the same results. We have also performed similar simulations for the PC with the basis shown in Fig. 1(b) with the same result.

\section{4. Waveguide effect in the lens with a small size along the PC axis}

We have considered so far only PC slabs that are infinite in the axial $z$-direction. In this section, we investigate the effects of termination of the slab in the $z$-direction. Namely, we are interested in how power in the $x$-direction varies as a function of the distance from the slab interface. We still keep our crystal infinite in the $y$-direction by using periodic boundary conditions. We are not interested now in the losses due to reflection of waves from the interface. Therefore we define the power $P$ as the surface integral
  \begin{equation} 
    P(x)=\int_Q S d a,
    \label{P}
  \end{equation}
where $S$ is Poynting's vector in the $x$-direction and the surface $Q$ is perpendicular to this direction and located at a distance $x$ from the left surface of the PC slab (see Fig. 9(b)). Calculating this integral as a function of $x$ {\em inside the PC} per unit length in $y$-direction one can study the leaking of energy due to diffraction in the $z$-direction. Note that we ignore small absorption of energy in pure silicon so that diffraction is the only mechanism that provides the $x$-dependence of $P$. The diffraction depends on what kind of material is above and below the PC slab in the $z$-direction. One side of the slab is always surrounded by air while the other side adjoins different media, namely air, silica or silicon.

\begin{figure}[htbp]
	\centering
	\includegraphics[width=8.6cm]{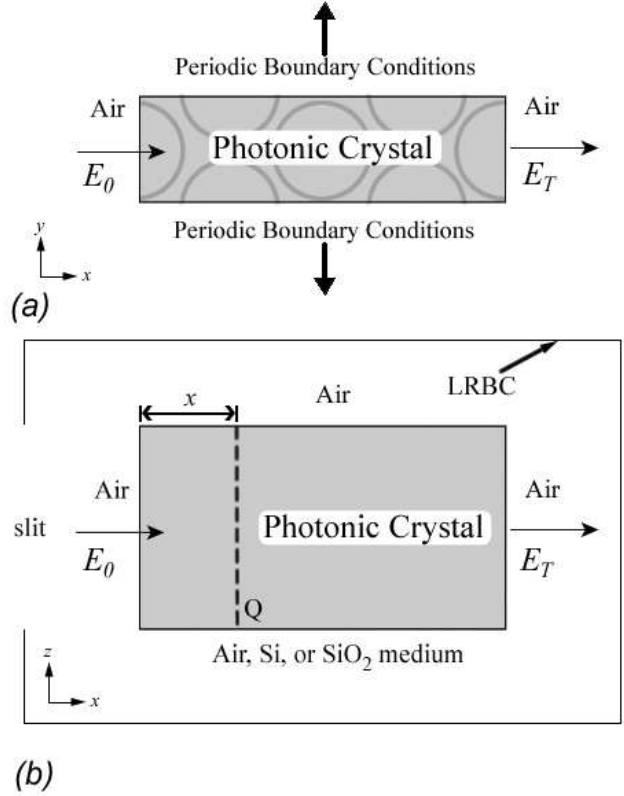}
	\caption{Representation of the PC slab geometry. (a) Top view of the cross-section of the PC in the $x-y$ plane. The system is infinite in $y$-direction due to periodic boundary conditions. (b) Side-view. Normal light is incident through a slit on the left while the remaining external boundaries are low-reflecting. Normal power flux is calculated across the surface $Q$ indicated by dashed line.
	\label{geom}}
\end{figure}

\begin{figure}[htbp]
	\centering
	\includegraphics[width=8.6cm]{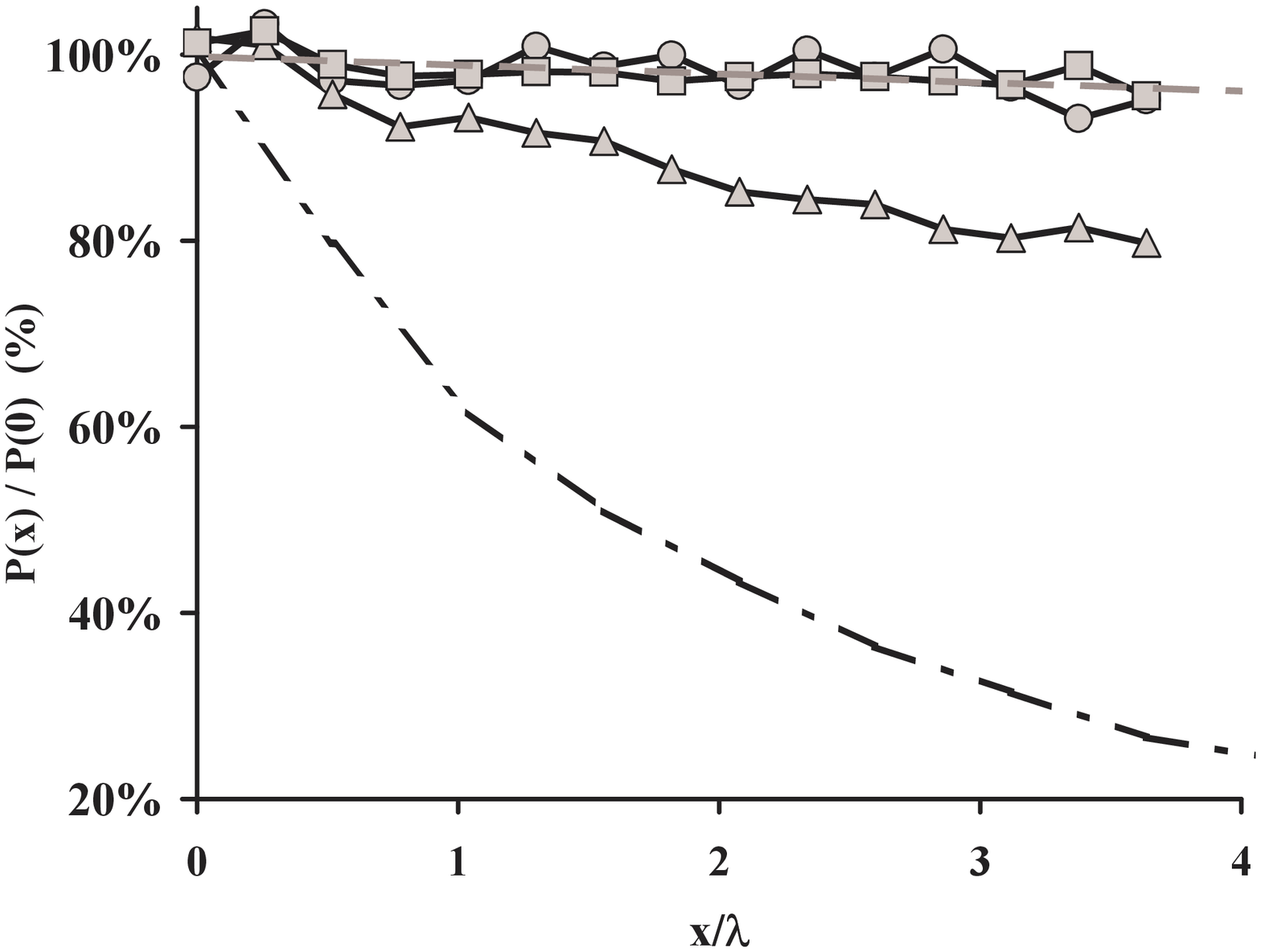}
	\caption{ $P(x)/P(0)$ in percent in the PC surrounded by air (solid line circles) and on substrates of silicon (solid line triangles) and silicon dioxide (solid line squares). The dash-dotted line shows the same power ratio when the PC slab is substituted by air and the ``substrate'' is also air. The trend line (dashed line) indicates least squares linear fit of the PC in air data. 
	\label{trans}}
\end{figure}

A 2D, top-view representation of our model is shown in the $x-y$ plane in Fig. 9(a). Periodic boundary conditions in the $y$-direction were used to simulate an infinite slab in this direction. As depicted in Fig. 9(b), an s-wave is incident normally through the slit in the boundary with low-reflecting boundary conditions (LRBC). The incident wave then goes through the air buffer to the PC slab surface. The transmitted and reflected field is almost perfectly absorbed by the surface with LRBC surrounding the system. We measure the power transmission in the $x$-direction as a function of distance in the slab.  The size of our model is limited by the computational memory requirements for large 3D geometries; we can accurately model a slab about seven lattice cells thick. We use AH cut of the surface 
 with  length in the $x$-direction  $ 3.54\lambda$ and with height in the $z$-direction $0.96 \lambda$.

The results of our simulation are shown in Fig. 10. One can see that the slab surrounded by air on all sides is a very good waveguide. The loss of electromagnetic power leaking from the slab in the $z$-direction of about 1.27\% per wavelength or 0.38\% per lattice cell ($\sqrt{3}d$). If, however, the slab were replaced by air, power flux transmission drops significantly faster due to  regular diffraction from the slit. Thus the PC slab suppresses diffraction. We think that the suppression is due to total internal reflection while all remaining losses are due to dipole radiation from the PC surface. A very low refractive index $n=1$ for the propagating waves is only because of the proximity of the wave vector with $k_z=0$ to the $\Gamma$-point. A two-dimensional band diagram shown in Fig. 2 does not have any meaning for a diffracted wave with $k_z\neq 0$. The diffracted waves, therefore, should have a much larger value of $n$ and should experience total internal reflection at the interface with air. To check this hypothesis we simulated the PC slab on a substrate made from silica or silicon. The silica substrate ($n=1.45$) does not prevent wave guiding, but the silicon substrate ($n=3.6$) makes it worse. The losses with the Si-substrate become  about 1.82\% per lattice cell. Power loss, however, is still much smaller than one would expect from diffraction; perhaps  there is an additional mechanism for wave guiding. Thus, both substrates are good candidates for the construction of a Veselago lens.

\section{5. Conclusion}
To summarize, we propose to make a Veselago lens using a uniaxial PC with a triangular lattice of cylindrical holes as a LHM. The working frequency is chosen in such a way that the refractive index for propagating modes $n=1$. Using air as a regular material one can match $n$, but cannot match the impedance. The lens, therefore, is multi-focal. The EW's decay inside the PC and  superlensing is absent. However, the half-width of the image is about $0.5\lambda$, in agreement with the diffraction limit for the flat lens. The additional foci are more smeared. Unexpectedly, the lens has good wave guiding properties that are mainly due to  total internal reflection, the quality of which depends on the substrate. 
\bibliography{evanh}

\begin{thebibliography}{13}
\expandafter\ifx\csname natexlab\endcsname\relax\def\natexlab#1{#1}\fi
\expandafter\ifx\csname bibnamefont\endcsname\relax
  \def\bibnamefont#1{#1}\fi
\expandafter\ifx\csname bibfnamefont\endcsname\relax
  \def\bibfnamefont#1{#1}\fi
\expandafter\ifx\csname citenamefont\endcsname\relax
  \def\citenamefont#1{#1}\fi
\expandafter\ifx\csname url\endcsname\relax
  \def\url#1{\texttt{#1}}\fi
\expandafter\ifx\csname urlprefix\endcsname\relax\def\urlprefix{URL }\fi
\providecommand{\bibinfo}[2]{#2}
\providecommand{\eprint}[2][]{\url{#2}}

\bibitem[{\citenamefont{Veselago}(1967)}]{ves}
\bibinfo{author}{\bibfnamefont{V.~G.} \bibnamefont{Veselago}},
  \bibinfo{journal}{Sov.\ Phys.-Solid\ State} \textbf{\bibinfo{volume}{8}},
  \bibinfo{pages}{2854} (\bibinfo{year}{1967}).

\bibitem[{\citenamefont{Efros and Pokrovsky}(2004)}]{ef}
\bibinfo{author}{\bibfnamefont{A.~L.} \bibnamefont{Efros}} \bibnamefont{and}
  \bibinfo{author}{\bibfnamefont{A.~L.} \bibnamefont{Pokrovsky}},
  \bibinfo{journal}{Solid\ State\ Comm.} \textbf{\bibinfo{volume}{129}},
  \bibinfo{pages}{643} (\bibinfo{year}{2004}).

\bibitem[{\citenamefont{Pokrovsky and Efros}(2002)}]{pok}
\bibinfo{author}{\bibfnamefont{A.~L.} \bibnamefont{Pokrovsky}}
  \bibnamefont{and} \bibinfo{author}{\bibfnamefont{A.~L.} \bibnamefont{Efros}},
  \bibinfo{journal}{Solid\ State\ Comm.} \textbf{\bibinfo{volume}{124}},
  \bibinfo{pages}{283} (\bibinfo{year}{2002}).

\bibitem[{\citenamefont{Parimi{\it \ et\ al.}}(2004)}]{parimis}
\bibinfo{author}{\bibfnamefont{P.~V.} \bibnamefont{Parimi{\it \ et\ al.}}},
  \bibinfo{journal}{Phys.\ Rev.\ Lett.} \textbf{\bibinfo{volume}{92}},
  \bibinfo{pages}{127401} (\bibinfo{year}{2004}).

\bibitem[{\citenamefont{A.~L.~Efros and Pokrovsky}()}]{cond1}
\bibinfo{author}{\bibfnamefont{C.~Y.~L.} \bibnamefont{A.~L.~Efros}}
  \bibnamefont{and} \bibinfo{author}{\bibfnamefont{A.~L.}
  \bibnamefont{Pokrovsky}}, \bibinfo{note}{cond-mat/0503494}.

\bibitem[{\citenamefont{C.~Y.~Li and Efros}()}]{cond2}
\bibinfo{author}{\bibfnamefont{J.~M.~H.} \bibnamefont{C.~Y.~Li}}
  \bibnamefont{and} \bibinfo{author}{\bibfnamefont{A.~L.} \bibnamefont{Efros}},
  \bibinfo{note}{cond-mat/0507090}.

\bibitem[{\citenamefont{Wang et~al.}(2004)\citenamefont{Wang, Ren, and
  Kempa}}]{w1}
\bibinfo{author}{\bibfnamefont{X.}~\bibnamefont{Wang}},
  \bibinfo{author}{\bibfnamefont{Z.~F.} \bibnamefont{Ren}}, \bibnamefont{and}
  \bibinfo{author}{\bibfnamefont{K.}~\bibnamefont{Kempa}},
  \bibinfo{journal}{Opt. Express} \textbf{\bibinfo{volume}{12}},
  \bibinfo{pages}{2919} (\bibinfo{year}{2004}).

\bibitem[{\citenamefont{Pokrovsky and Efros}(2003)}]{poef}
\bibinfo{author}{\bibfnamefont{A.~L.} \bibnamefont{Pokrovsky}}
  \bibnamefont{and} \bibinfo{author}{\bibfnamefont{A.~L.} \bibnamefont{Efros}},
  \bibinfo{journal}{Appl. Opt.} \textbf{\bibinfo{volume}{42}},
  \bibinfo{pages}{5701} (\bibinfo{year}{2003}).

\bibitem[{\citenamefont{Wang et~al.}(2005)\citenamefont{Wang, Ren, and
  Kempa}}]{w2}
\bibinfo{author}{\bibfnamefont{X.}~\bibnamefont{Wang}},
  \bibinfo{author}{\bibfnamefont{Z.~F.} \bibnamefont{Ren}}, \bibnamefont{and}
  \bibinfo{author}{\bibfnamefont{K.}~\bibnamefont{Kempa}},
  \bibinfo{journal}{Appl. Phys.\ Lett.} \textbf{\bibinfo{volume}{86}},
  \bibinfo{pages}{061105} (\bibinfo{year}{2005}).

\bibitem[{\citenamefont{Jackson}(1998)}]{jac}
\bibinfo{author}{\bibfnamefont{J.~D.} \bibnamefont{Jackson}},
  \emph{\bibinfo{title}{Classical Electrodynamics}} (\bibinfo{publisher}{Willy
  and Sons, New York}, \bibinfo{year}{1998}).

\bibitem[{\citenamefont{Pendry}(2000)}]{pen}
\bibinfo{author}{\bibfnamefont{J.~B.} \bibnamefont{Pendry}},
  \bibinfo{journal}{Phys.\ Rev.\ Lett.} \textbf{\bibinfo{volume}{85}},
  \bibinfo{pages}{3966} (\bibinfo{year}{2000}).

\bibitem[{\citenamefont{Haldane}()}]{hal}
\bibinfo{author}{\bibfnamefont{F.~D.~M.} \bibnamefont{Haldane}},
  \bibinfo{note}{cond-mat/0206420}.

\bibitem[{\citenamefont{Ruppin}(2002)}]{rup}
\bibinfo{author}{\bibfnamefont{R.}~\bibnamefont{Ruppin}},
  \bibinfo{journal}{Phys.\ Lett. A} \textbf{\bibinfo{volume}{229}},
  \bibinfo{pages}{309 } (\bibinfo{year}{2002}).

\end{thebibliography}
\end{document}